\RequirePackage[mathlines]{lineno}
\documentclass[aps, twocolumn, superscriptaddress, bibnotes]{revtex4}

\usepackage{amsfonts}           
\usepackage{amssymb}            
\usepackage{amsmath}            
\usepackage{comment}
\usepackage{latexsym}           
\usepackage{dcolumn}            
\usepackage[dvips]{graphicx}   
\usepackage{enumerate}
\usepackage{epstopdf}           
\usepackage{fancyhdr}           
\usepackage{hyperref}           
\usepackage{setspace}           
\usepackage{xspace}             
\usepackage{subfigure}          
\usepackage{bm}                 
\usepackage[ulem=normalem]{changes}
\usepackage{url}
\usepackage{color}
\newcommand{\bea}{\begin{eqnarray}}
\newcommand{\eea}{\end{eqnarray}}

\newcommand{\eg}{{\it e.g.}}

\begin{document}
\title{Probing microplasticity in small scale FCC crystals via Dynamic Mechanical Analysis}
\date{\today}

\author{Xiaoyue Ni}
\email[To whom correspondence should be addressed:\\]{xnni@caltech.edu}
\affiliation{Division of Engineering and Applied Sciences, California Institute of Technology, Pasadena, CA 91125}

\author{Stefanos Papanikolaou}
\affiliation{The Johns Hopkins University, Hopkins Extreme Materials Institute, Baltimore, MD 21218}

\author{Gabriele Vajente}
\affiliation{LIGO Laboratory, California Institute of Technology, Pasadena, CA 91125}

\author{Rana X Adhikari}
\affiliation{LIGO Laboratory, California Institute of Technology, Pasadena, CA 91125}

\author{Julia R. Greer}
\affiliation{Division of Engineering and Applied Sciences, California Institute of Technology, Pasadena, CA 91125}
\begin{abstract}
In small-scale metallic systems, collective dislocation activity has been correlated with size effects in strength and with a step-like plastic response under uniaxial compression and tension. Yielding and plastic flow in these samples is often accompanied by the emergence of multiple dislocation avalanches. Dislocations might be active pre-yield, but their activity typically cannot be discerned because of the inherent instrumental noise in detecting equipment. We apply Alternate Current (AC) load perturbations via Dynamic Mechanical Analysis (DMA) during quasi-static uniaxial compression experiments on single crystalline Cu nano-pillars with diameters of 500 nm, and compute dynamic moduli at frequencies 0.1, 0.3, 1, and 10 Hz under progressively higher static loads until yielding. By tracking the collective aspects of the oscillatory stress-strain-time series in multiple samples, we observe an evolving dissipative component of the dislocation network response that signifies the transition from elastic behavior to dislocation avalanches in the globally pre-yield regime. We postulate that microplasticity, which is associated with the combination of dislocation avalanches and slow viscoplastic relaxations, is the cause of the dependency of dynamic modulus on the driving rate and the quasi-static stress. We construct a continuum mesoscopic dislocation dynamics model to compute the frequency response of stress over strain and obtain a consistent agreement with experimental observations. The results of our experiments and simulations present a pathway to discern and quantify correlated dislocation activity in the pre-yield regime of deforming crystals. 
\end{abstract}

\maketitle

\begin{figure}[t!]
\centering
\includegraphics[scale=0.6]{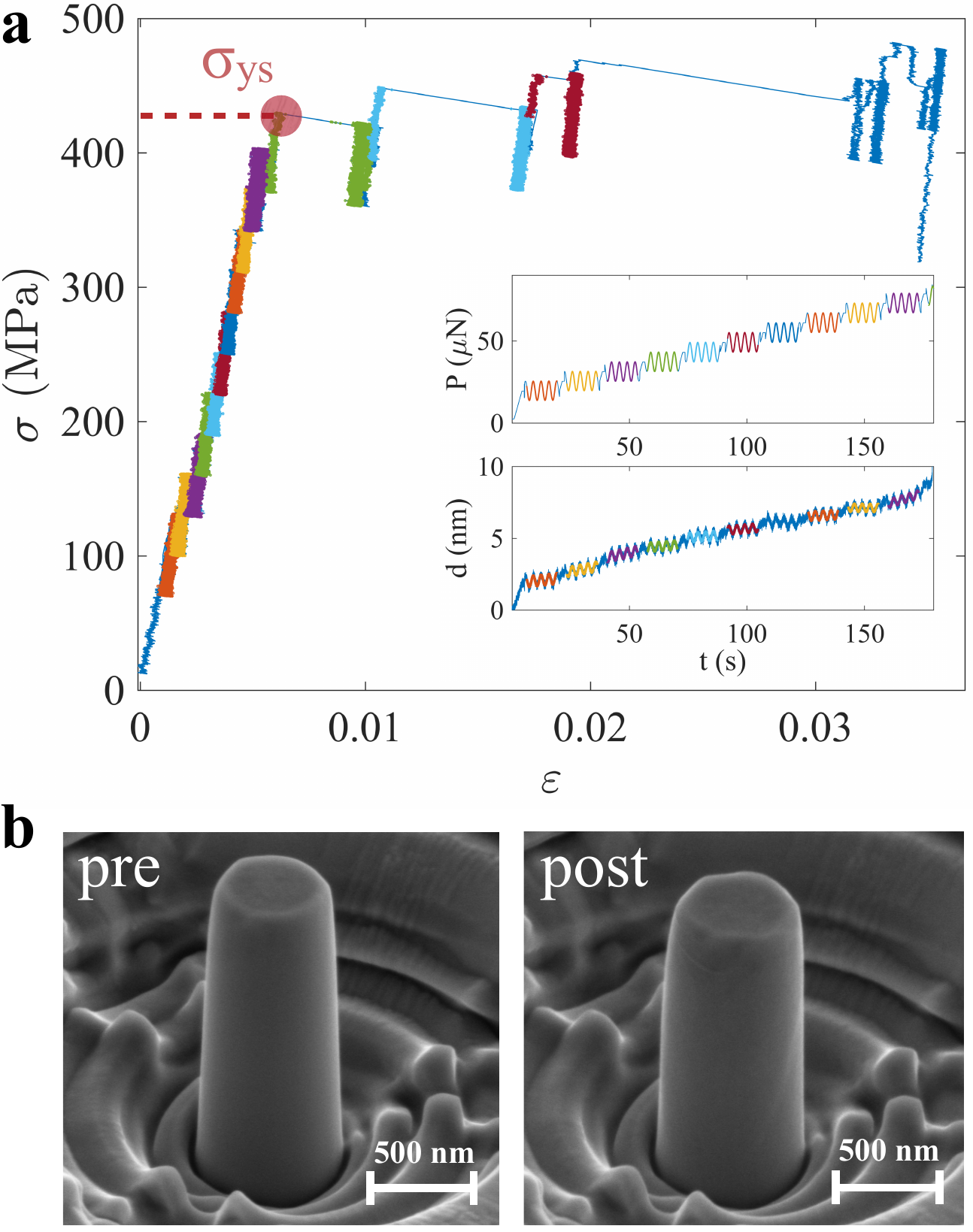}
\caption{{\bf Dynamic Mechanical Analysis on Cu nano-pillars.} (a) Engineering stress vs. strain during DMA measurements on a Cu sample at a frequency of 0.3 Hz; different colors separate data taken at different steps of stress-oscillation segments. The inset shows raw load vs. displacement as a function of time in the pre-yield regime. The overlaying different color curves are sinusoidal fits for each data set of oscillation segments, using Eqs. \ref{eq:fita} and \ref{eq:fitb}. (b) SEM images of an as-fabricated (pre-) and compressed (post-) $\sim 500~nm$ diameter Cu pillar with a nominal aspect ratio of 3:1.}
\label{fig:1}
\end{figure}

{\it Introduction.--}Mechanical deformation of materials is usually described by continuous, deterministic stress-strain relations, for examples see Ref.~\cite{asaro}. In the last decade, Uchic {\it et~al.} first applied the uni-axial compression methodology on focused ion beam (FIB)-machined Ni micro-pillars \cite{Uchic:2004}. Greer and Nix then extended it to Au nano-pillars \cite{Greer:2005}, and since then the discrete and stochastic deformation of small-scale single-crystalline metals has been ubiquitously observed \cite{Dimiduk:2006, Uchic:2009, Greer:2011}. The large strain bursts are unambiguously distinguished as serrations in the stress-strain curves as shown in Fig.~\ref{fig:1}(a). They have been mainly attributed to the unique nano-scale plasticity mechanisms, where the operation of individual dislocation sources, single-arm or surface, governs deformation and strength \cite{TAP:2007, Rao:2007}. The extent of these strain bursts usually ranges from nanometers to a few microns \cite{Uchic:2004, Dimiduk:2006, Friedman:2012, Maass:2013}. The analysis of strain bursts shows that the slip size distributions follow power laws \cite{Dimiduk:2006, Sethna:2001}, with system-size- \cite{Csikor:2007, Brinckmann:2008} and stress- \cite{Dahmen:2009, Friedman:2012} dependent cutoffs. It is unclear whether smaller strain bursts, undetected by the instrument, are present in the deformation of such micro- and nano-sized single crystals, especially prior to the yield point, commonly defined as the start of the first detected burst, as shown in Fig.~\ref{fig:1}(a) at stress $\sigma_{ys}$. While small, these events compose mechanical noise that is imperative to understand and remove for high-precision experiments, such as the Advanced Laser Interferometer Gravitational-Wave Observatory (LIGO) \cite{LIGO, Vajente:2016} -- the impulsive strain events propagated from the metallic suspension system to the test mass \cite{suspension} can introduce background noise which could limit the interferometer sensitivity.
 
Using basic forms of mechanical loading, such as a force- or displacement-controlled compression, careful examination of the data provided evidence for the presence of short plastic instabilities before the onset of the obvious and apparent strain bursts \cite{Uchic:2004, Greer:2005, Jennings:2010, Jennings:2011, Friedman:2012, Maass:2015} (\eg~100 - 400 MPa regime of 500 nm pillars, as shown in Fig.~\ref{fig:1}(a)). In-situ transmission electron microscope (TEM) nanoindentation experiments revealed the onset of dislocation motion before the first obvious displacement excursion \cite{Minor:2006, Shan:2007}. Creep experiments on single crystals of ice detected acoustic emission events at resolved shear stresses far below the yield stress \cite{Miguel:2001}. These observations have yet to be connected to constitutive relations and a quantifiable stress-strain response. Discrete Dislocation Dynamics (DDD) simulations suggest the existence of intermittent events in the pre-yield regime of crystalline materials \cite{Laurson:2012, Ispanovity:2014} and a significant loading rate effect on strain burst response of nano- and micro-crystals due to dislocation jamming and relaxation \cite{Cui:2016}. Stress-induced probabilistic cross-slip relaxation has also been associated with several non-trivial aspects of crystal plasticity \cite{Papanikolaou:2012}. It's natural to question whether we can detect and quantify microplasticity in crystals' pre-yield regime.  

Machine noise has been the Achilles' heel of numerous experimental nano-mechanical investigations. Attempts have been made to characterize the machine noise, with reported values of $\sim$ 0.2 nm displacement-, $\sim$ 30 nN force- noise floor, and a thermal drift of $ < 0.05$ nm/s for the prevalently used Hysitron TI 950 Triboindenter in quasi-static mode \cite{Hysitron}. In uniaxial compression experiments, a flat nanoindenter tip applies compressive load to the top of a commonly cylindrical sample, a so-called “micro- or nano-pillar”, and the indenter-sample friction, as well as the electromagnetic assembly responsible for the load control produce substantial and inevitable machine noise. In addition, noise caused by thermal drift sets a limit on the duration of such experiments, which renders long-time mechanical experiments like cyclical or fatigue loading, as well as creep tests, virtually impossible to interpret. Statistical probing is necessary to detect any possible non-linear dislocation activities, which cause axial displacements below the machine noise. We apply Dynamic Mechanical Analysis (DMA) at multiple frequencies that span three orders of magnitude, from 0.1 to 10 Hz, on multiple 500 nm-diameter single crystalline Cu nano-pillars. We statistically characterize the overall DMA behavior and compare it with mean-field dislocation depinning predictions.

{\it Experiment.--}Cylindrical nano-pillars with diameters of $\sim 500$ nm and aspect ratios (height/diameter) of $\sim$ 3:1 were fabricated following a concentric-circles top-down methodology using a Focused Ion Beam \cite{Brinckmann:2008, Kim:2009, Kim:2010} from bulk single-crystalline copper ($ > 99.9999 \%$ purity) with one side polished to a $< 30~\mathring{A}$ RMS roughness, oriented in $\sim \langle111\rangle$ direction \cite{MTI}. The nano-mechanical experiments were carried out in a nanoindenter (Triboindenter, Hysitron \cite{Hysitron}) equipped with an 8 $\mu$m-diameter flat punch diamond tip custom made specifically for these experiments. Fig.~\ref{fig:1}(a) conveys a representative compressive engineering stress-strain data, with the inset showing the corresponding time series of load and displacement, zoomed into the pre-yield regime. In the experiment, we applied a uniaxial quasi-static load that monotonically increased in a step-wise fashion to an individual nano-pillar. Small stress oscillations with the amplitude of $6~\mu$N and a fixed frequency in the range between 0.1 and 10 Hz were superimposed over the static load to each 15~s step interval. Before the initiation of each compression experiment we waited for $>$ 145 s to equilibrate the in-contact displacement drift and used the last 20~s drift data to estimate the thermal drift rate for subsequent correction. Only those experiments where the thermal drift rate was less than 0.05~nm/s were analyzed. The loading time before the occurrence of the first large strain event was usually within the first 200~s for all tests. Fig.~\ref{fig:1}(b) shows the representative pre- and post-compression SEM images of a representative Cu nano-pillar.  

The dynamic modulus is defined as the frequency response of stress over strain $E(\omega, \sigma_0) = \frac{\sigma(\omega)}{\varepsilon(\omega)}$, where $\sigma_0$ is the applied quasi-static stress and $\omega = 2\pi f$ is the driving frequency. Using this definition, we can extract the dynamic modulus from the oscillations that are imposed at each quasi-static stress $\sigma_0$ using a frequency domain analysis. We fit the time series of stress $\sigma(t)$ and strain $\varepsilon(t)$ using the following form which also includes a linear drift term:
\begin{subequations}
	\begin{align}
		\sigma_f(t) &= x_r cos(\omega t)+x_i sin(\omega t)+\sigma_d t + \sigma_0, \label{eq:fita}\\
		\varepsilon_f(t) &= u_r cos(\omega t)+ u_i sin(\omega t)+\varepsilon_d t + \varepsilon_0, \label{eq:fitb}
	\end{align}
\end{subequations}
where $x_r, x_i, \sigma_d, \sigma_0, u_r, u_i, \varepsilon_d, \varepsilon_0$ are fitting parameters for the stress and strain. The complex dynamic modulus $E$ can then be calculated as a function of the $\omega$ and $\sigma_0$:
\begin{equation}
	E(\omega, \sigma_0) = \frac{x_r-i x_i}{u_r -i u_i} = A(\omega, \sigma_0) e^{i \phi(\omega, \sigma_0)}, \label{eq:dynmod}
\end{equation}
where $A$ and $\phi$ are the amplitude and phase components of the dynamic modulus.

\begin{figure}[t!]
\centering
\includegraphics[scale=0.68]{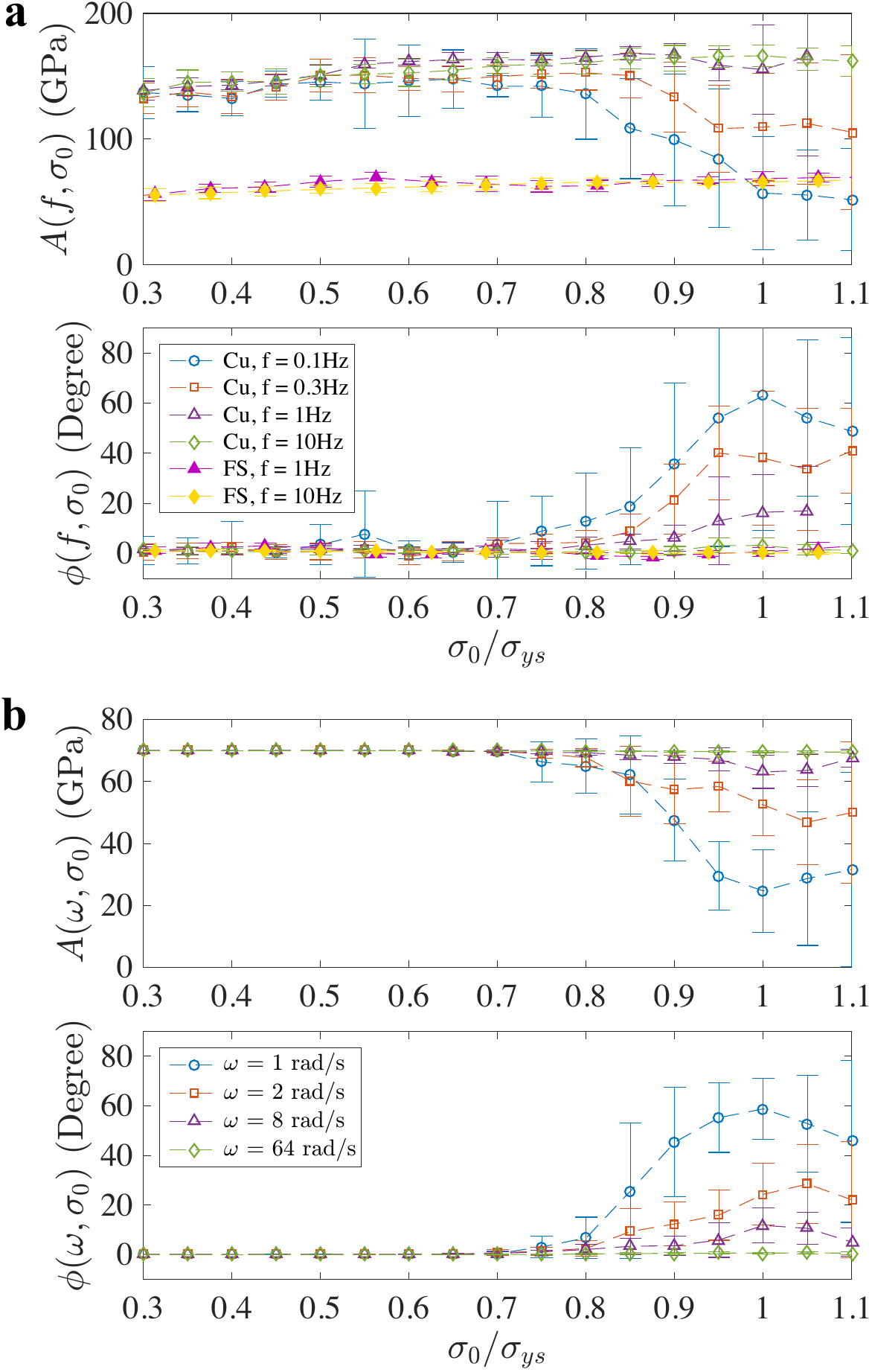}
\caption{(a) {\bf Dynamic Modulus vs. Stress.} The experimental DMA results on copper and fused silica (FS) 500 nm pillars. The amplitude A and phase $\phi$ of the dynamic modulus values are statistically analyzed from 6 samples for driving varying from 0.1 Hz to 10 Hz, both plotted versus the global stress, around which the oscillations were applied. This plot demonstrates that the dynamic modulus is not constant as the quasi-static load approaches the yield point, and the deviation gets larger with slower driving. (b) DMA results from the mesoscopic dislocation dynamics simulation. }
\label{fig:2}
\end{figure}

We applied this type of DMA using different driving frequencies: 0.1, 0.3, 1, and 10 Hz, for each quasi-static load hold and took measurements from 6 samples for each frequency driving test. We solved for the dynamic modulus at each quasi-static loading step at the single driving frequency using the fitting procedure described above. The quasi-static stress at each step is normalized by the yield stress of the system $\sigma_{ys}$. We binned amplitude and phase lag for sample statistics. The binning mean and standard error for amplitude and phase lag were calculated as a function of the stress bin centers and are shown for each driving frequency in Fig.~\ref{fig:2}(a). This DMA data reveals a maximum of $\sim 70\%$ decrease in the average amplitude and a maximum of $\sim 60^\circ$ increase in the average phase lag as the applied quasi-static stress approaches yielding at 1 ($\sim$ 400 MPa). This plot also shows that these deviation from elastic behavior are more pronounced for slower driving frequencies. These results are in stark contrast to the DMA data collected from the same type of uniaxial compression on a $\sim$ 500nm-diameter fused silica nano-pillars, which exhibits a constant amplitude of $\sim 65$ GPa and a no-delay response for the driving frequencies of 1 Hz and 10 Hz.

{\it Simulation.--}To reveal the underlying mechanisms that drive the observed non-trivial loss behavior in Cu as the applied stress approaches yielding, we constructed a continuum crystal plasticity model that aims at capturing the salient aspects of the observed mechanical behavior. This model considers the energetics of two competing processes: the dislocation-driven abrupt strain jumps and the slow stress-controlled relaxations towards minimum system energy state. To capture both the fast avalanches and the slow viscoplastic relaxations, we utilized a cellular automaton constitutive microplasticity model enhanced with an additional continuous-in-time strain field that follows a viscoplastic constitutive law \cite{Zaiser:2006, Papanikolaou:2012}. We model the shear strain to consist of the elastic and plastic components $\gamma = \gamma_e + \gamma_p$. The elastic term is calculated using Hooke's law. The plasticity model that captures the plastic term can be realized using detailed continuum plasticity modeling approaches \cite{asaro}. It is reasonable to assume that in a single representative volume element for single-crystalline FCC crystals, the following criteria hold: i) uniaxial loading activates one dominant crystallographic slip system, A, with another system, B, assisting dislocation glide along A \footnote{Although the bulk is nominally high-symmetry orientated, in a large deformation picture the pillars would point towards dominant slip systems. The $\sim \langle111\rangle$ orientation leads to a slip-system with near zero resolved shear stress, leaving lots of dislocations that can function as the B slip system \cite{Jennings:2010}.} and ii) dislocations carry plastic distortion via two distinct mechanisms: (a) fast dislocation avalanche-like glide and (b) slow, stress-relaxation-driven secondary glide on A caused by the coupled A-B dislocation mechanisms (\eg~double cross-slip) \cite{Papanikolaou:2012}.  With contributions from both mechanisms, the total plastic strain can be expressed as:
\begin{equation}
	\gamma_p  = \gamma_p^{(a)}+ \gamma_p^{(b)}.\label{eq:motion}
\end{equation}
In the fast dislocation avalanche-driven mechanism, a volume element at location $\mathbf{r}$ yields a random plastic strain $\delta \gamma_p^{(a)}$ if the local stress $\tau(\mathbf{r})$ is larger than a local depinning threshold $\chi(\mathbf{r})$ \cite{fisher, Papanikolaou:2012, Numeric:1988}, where $\chi(\mathbf{r})$ follows a uniform distribution \cite{supplement}. After each avalanche, the threshold value is re-drawn from the same distribution. 
On the other hand, the slow relaxation mechanism follows a typical constitutive viscoplastic law: 
\begin{equation}
	\dot{\gamma}_p^{(b)} = \frac{D}{G} (\tau(\mathbf{r}))^{n},
\end{equation}
where $D$ is the relaxation constant, $G$ is the shear modulus, and $n\in[1,3] < 10$ is the critical quantity to define another timescale which is slow compared to the fast avalanche process \cite{Jennings:2011}.

For numerical simplicity, we apply this methodology to edge dislocations only, for which the local resolved shear stress can be explicitly calculated:
\begin{equation}
\begin{aligned}
\tau(\mathbf{r}) &= \tau_{ext} + \tau_{int}(\mathbf{r}) + \tau_{hard}(\mathbf{r}) \\
&= \tau_0 + \tau_A sin(\omega t) + \int d^2\mathbf{r}'K(\mathbf{r}-\mathbf{r}')\gamma_p(\mathbf{r}') - h\gamma_p(\mathbf{r}), \label{eq:stress}
\end{aligned}
\end{equation}
where $\tau_{ext}$ is the applied external quasi-static stress combined with the oscillation component, $\tau_{int}$ is the stress that accounts for the long-range interactions with other dislocations, and $\tau_{hard}$ is the stress that arises from dislocation hardening. In the expanded form of Eq. \ref{eq:stress}, $K$ serves as the interaction kernel for single slip straight edge dislocations, and $h$ represents a mean-field phenomenological hardening parameter \cite{Zaiser:2006}. For the stress kernels of complete circular dislocation loops or screws, in principle the results would be unchanged, since all these kernels are sufficiently long-ranged \cite{fisher, Ghoniem:2002}. The model implementation is such that the system is meshed into $N \times N$ elements, with $N$ = 32. We prescribe similar loading conditions to 8 random initial configurations as we did in the experiments, with different driving frequencies of 1, 2, 8, and 64 rad/s. The rate equation associated to Eq. \ref{eq:motion} can be numerically solved by Euler integration with a fixed time step $\Delta t = 10^{-2} s$. Additional simulation details are provided in Supplemental Materials \cite{supplement}.

{\it Discussion.--}Fig.~\ref{fig:2}(b) shows the same frequency domain analysis of the dynamic modulus using simulations results as the ones shown in Fig.~\ref{fig:2}(a) for the experimental data. This qualitative agreement between simulations and experiments motivates a further quantitative comparison. Existing simulations investigated the effect of cyclic loading on the evolved dislocation network and predicted a scaling relation between the normalized strain rate amplitude and the driving frequency, focused on the mean-field depinning theory framework \cite{Schutze:2011, Laurson:2012}:
\begin{equation}
\frac{|\dot{\varepsilon}|}{|\sigma|} \sim \omega^\kappa \label{powerlaw}
\end{equation}
\begin{figure}[t!]
\centering
\includegraphics[scale=0.44]{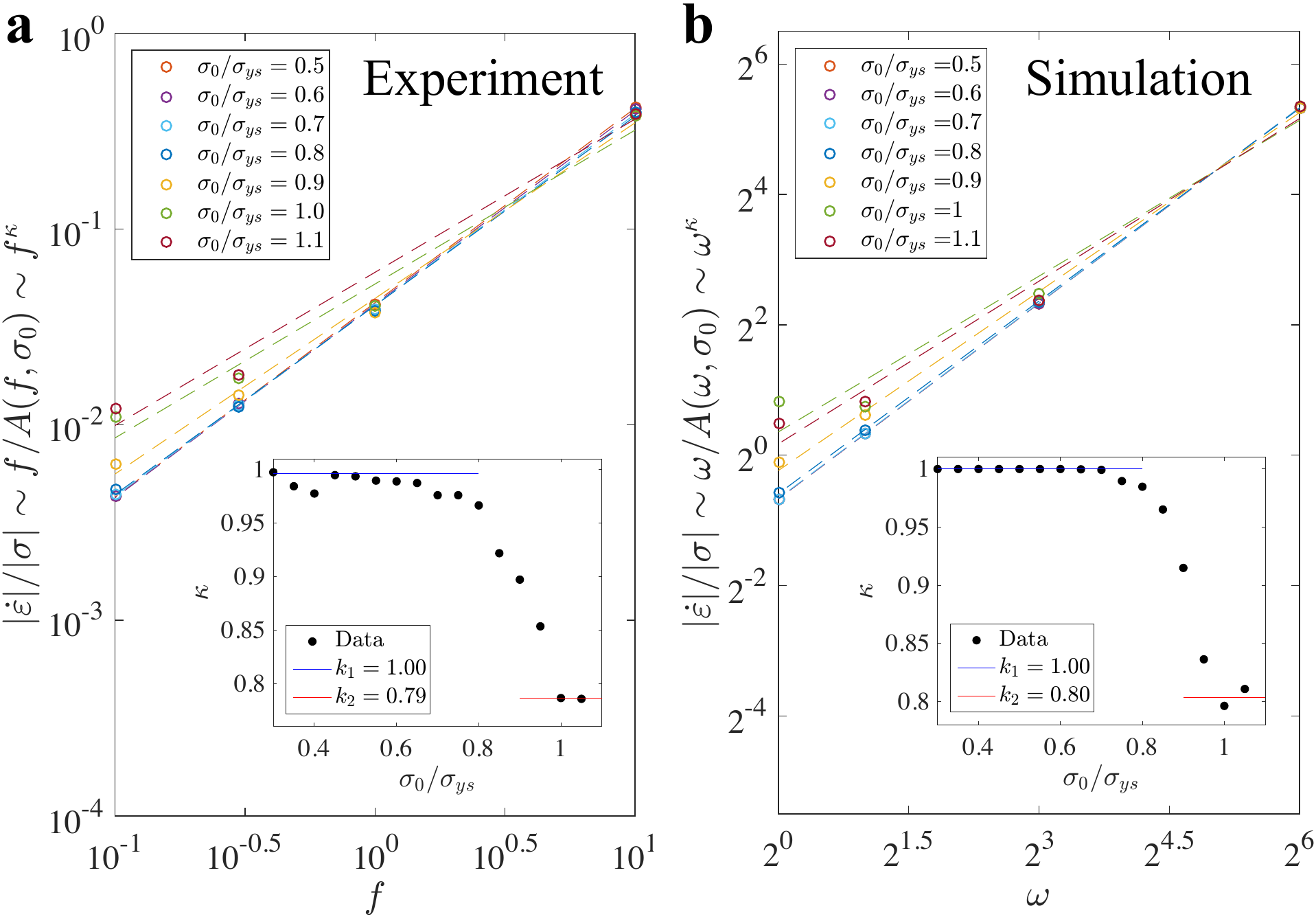}
\caption{{\bf Scaling Analysis of the Normalized Strain-rate Amplitude.} (a) The normalized strain rate amplitude scaling over driving frequency analysis \cite{Laurson:2012} using experimental DMA data, and (b) simulation DMA data. The figures show explicitly the fitting for scaling parameter $\kappa$ using Eq. \ref{powerlaw} at different quasi-static stress levels. The inset presents the measured $\kappa$ as a function of normalized stress.}
\label{fig:3}
\end{figure}
where $\kappa$ = 1 corresponds to a simple harmonic oscillator, i.e. perfectly elastic behavior, and $\kappa$ = 0.82 corresponds to a system driven close to the pinning threshold \cite{Schutze:2011, Laurson:2012}.  The strain rate amplitude is normalized by the stress amplitude, $|\dot{\varepsilon}|/|\sigma|$ which is equivalent to $\omega/A$, where $A$ is the dynamic modulus amplitude measured via DMA. Fig.~\ref{fig:3} shows the scaling analysis of the normalized strain rate amplitude vs. driving frequency for the dynamic modulus amplitude calculated from the experiments and simulations at different quasi-static loads. The insets show the scaling parameter $\kappa$ as a function of the normalized stress. These plots convey that at both small and large stress regimes, experiments and simulations produce scaling behaviors that are in agreement with the mean-field depinning predictions, and a smooth, microplastic crossover connects these two extreme regimes. The experiments and simulations reveal enhanced microplasticity activities as the system is stressed close to yielding. The actual mechanism that is responsible for the increased `susceptibility' to plasticity can be a thermally activation process like cross-slip, or the collective dislocation bowing out due to long-range interactions, {\it i.e.} the Andrade mechanism \cite{Miguel:2002}.

{\it Summary.--}We imposed oscillatory loads in the nominal elastic regime of the uniaxially compressed 500 nm-diameter single crystalline Cu nano-pillars. We applied monotonically increasing stresses above the bulk yield point of 70 MPa \cite{Copper} to investigate the mechanically correlated material response. Analysis of the cumulative oscillatory response reveals a substantial deviation from the nominally perfectly elastic behavior, as well as an emergent dissipation signature in what has always been considered pre-yield regime. Our experimental observations are corroborated by a mesoscale dislocation plasticity model, which accounts for dislocation avalanches (fast processes) and the viscoplastic response (slow time scales) during oscillatory loading. We formulate a scaling analysis that shows a smooth transition of the system from perfect elasticity to dislocation depinning-driven plasticity that occurs at loads lower than the global yield stress. This approach represents a new pathway to investigate and quantify the abrupt plastic events that emanate from dislocation activities even in the pre-yield regime, that occur ubiquitously during deformation of small-scale single crystals below instrumental noise levels. 

The developed methodology can be applied to characterize pre-yield dislocation dynamics in extensive list of FCC, BCC, and HCP materials. The micromechanical study sheds light on detecting crackling noise in macroscopic sample subjected to nominal elastic loading. The observation of such events might lead to better prediction of plastic yielding and even incipient fracture for structural materials. The pre-yield mechanical noise itself can be a hidden problem for instrumentation that requires high strain sensitivity, {\it e.g.} advanced LIGO \cite{LIGO, Vajente:2016}. 

\begin{acknowledgements}
We would like to thank Eric K. Gustafson, Koji Arai, Eric Quintero, and the members of the Seismic Isolation and Suspension Working Group in the LIGO Scientific Collaboration for many enriching discussions. Also many thanks to Ottman Turtuliano for assistance in fused silica sample preparation. We gratefully acknowledge financial support from NSF Award No. DMR-1204864 (J. R. G.). LIGO was constructed by the California Institute of Technology and Massachusetts Institute of Technology with funding from the National Science Foundation, and operates under cooperative agreement PHY-0757058. Advanced LIGO was built under Award No. PHY-0823459. This paper carries LIGO Document No. LIGO-P1600316. This manuscript has an internal LIGO project designation of P1600316.
\end{acknowledgements}


\end{document}